\documentstyle[aps,preprint]{revtex}
\begin{document}

\title{Oscillatory screening of the dc electric field in the Si-SiO$_2$ multiple quantum wells probed by
second-harmonic generation}
\author{V. V. Savkin, A. A. Fedyanin, A. N. Rubtsov, and O. A. Aktsipetrov}
\address{Department of Physics, Moscow State University, 119899 Moscow,
Russia}
\author{F. A. Pudonin}
\address{ Lebedev Physical Institute of RAS, Leninsky pr. 53, 117924
Moscow, Russia}

\date{\today}

\maketitle

\begin{abstract}
DC-electric field, being screened in 3D semiconductors, normally decays monotonically in space.
Experimental studies of the DC electric field screening in Si-SiO$_2$ multiple quantum wells by electric field
induced optical second-harmonic generation show a non-monotonic, oscillatory-like decay. The model of
electrons localized inside quantum wells, with the first subband occupied, allows a description of the phenomenon.
Interwell Coulomb interaction, a finite value of the electron charge and strong effective-mass anisotropy result in a
crucial difference from 3D Fermi liquid.
\end{abstract}


\narrowtext
\par
Properties of the degenerate electron subsystem of a solid are basically determined by an interplay between
Coulomb interaction and quantum effects, which is revealed in correlation and exchange interaction.
The corresponding typical energy scales are given by characteristic magnitude of the Coulomb interaction
$\varepsilon _C\propto e^2n^{-1/3}\in ^{-1}$ and the Fermi energy for noninteracting electrons
$\varepsilon _F\propto \hbar ^2m_0^{-1}n^{-2/3}$, where
$n$ is the electron density, $\in $ denotes the lattice dielectric constant, $m_0$, $e$ and $\hbar $
are fundamental constants. If $\varepsilon _F/\varepsilon _C>>10^{-2}\div 10^{-1}$ the system has the
properties of a normal Fermi liquid\cite{1}, whereas in the opposite case a dielectric bandgap forms\cite{2}.
An Anderson-Mott transition occurs at the specific value of $\varepsilon _F/\varepsilon _C$ \cite{3}

\par    Significant attention has been paid, over recent decades, to spatially confined semiconductor-based
systems. Advanced modern technologies allow the fabrication of structures with a predetermined morphology,
and therefore control over their electronic properties. Amongst 2D semiconductor-based
systems (superlattices and multiple quantum wells) two extreme cases should be pointed out. While epitaxial
structures like GaAs/AlGaAs are characterized by relatively close bandgaps of the
components, the bandgap difference in amorphous Si-SiO$_2$ based systems are much higher than the Si
bandgap.

\par    The anisotropic dispersion law of electrons requires the consideration of two
parameters, $\left( {\varepsilon _F/\varepsilon _C}
\right)_{|\,|}$ and $\left( {\varepsilon _F/\varepsilon _C}
\right)_\bot $, which describe tangential and transversal electron
motion, respectively. For the Si layer with a carrier density of
$10^{16} \div 10^{18}\,cm^{-3}$, at the zero temperature and with
$m \approx m_0$ we have $\left( {\varepsilon _F/\varepsilon _C}
\right)_{|\,|}\approx 10^{-2}\div 10^{-1}$, i.e. lateral
properties of the system can be described in terms of a
Fermi-liquid. The parameter $\left( {\varepsilon _F/\varepsilon
_C} \right)_\bot $ is governed by the transparency of potential
barriers in the superstructure. For high potential barriers  the
parameter $\left( {\varepsilon _F/\varepsilon _C} \right)_\bot $
is very small, and the interlayer electron motion is suppressed by
the Coulomb repulsion. Electrons are localized within the layers
and the structure can be treated as a set of independent quantum
wells (QW), {\it i.e.} as multiple QW's (MQW). The Mott-like
transition is expected
 at a certain transparency of the barrier.

\par    Properties of the electron liquid reveal themselves in the response of the system to external
fields. For example, Friedel oscillations appear in the screening
cloud in the degenerate Fermi-liquid\cite{4}. The period of these
oscillations is governed by the typical de-Broiglie wavelength in
the system. 0D systems weakly interacting with their environment
show a Coulomb blockade phenomenon, which is due to a finite value
of the elementary charge\cite{5}. Analysis of four coupled quantum
dots\cite{6} shows that properties of an array of interacting
quantum dots depend on the interdot tunnel coupling. For weak
coupling this system behaves as a set of independent dots, at
intermediate coupling the Mott insulator forms (the so-called
 collective Coulomb blockade) and for strong coupling the Coulomb blockade is completely destroyed.
It is worth noting that for the collective Coulomb blockade the screening has an essentially collective nature,
but it still shows non-monotonic features. Particularly, the carrier density in neighboring quantum dots is
{\it anti-correlated}.

\par    On the other hand, the DC-electric-field screening in bulk semiconductors shows monotonic spatial
distribution of the voltage drop across the space charge
region\cite{7}. We conclude that nonmonotonic peculiarities in the
DC-screening indicate that the electron subsystem essentially
differs from the bulk 3D case.

\par    Optical second-harmonic generation (SHG) is a
versatile optical probe of surfaces and interfaces\cite{8}. This
technique is applicable to all interfaces accessible by light,
including buried interfaces\cite{9}, and has inherently high
spatial\cite{10}, spectral\cite{11} and temporal\cite{12}
resolutions. The sensitivity of SHG arises from the strong
symmetric selection rules for the second-order susceptibility.
Namely, the second-order susceptibility vanishes in the dipole
approximation for media with inversion symmetry, and the
quadrupole term becomes prevalent. In the presence of the DC-
electric field the dipole contribution into the second-order
polarization ${\bf P}^{2\omega }$ arises\cite{13}:

\begin{equation}
P^{2\omega }=i\chi ^{(2Q)}\vdots k^\omega E^\omega E^\omega  +\chi ^{(3D)}\vdots E^{DC} E^\omega
E^\omega ,
\label{eq1}
\end{equation}
where $E^\omega $ and $E^{DC}$ are the fundamental and DC fields,
respectively; $k^\omega $ is the fundamental wavevector; nonlinear
susceptibilities $\chi ^{(2Q)}$ and $\chi ^{(3D)}$  are
responsible for the $2\omega =\omega +\omega $ and $2\omega
=0+\omega +\omega$ processes, respectively. Two terms in
(\ref{eq1}) are comparable at $E^{DC}\approx10^5$ V/cm. This
sensitivity to the DC-fields has been experimentally demonstrated
for semiconductor-electrolyte and semiconductor-dielectric
interfaces\cite{13}.

\par    In this paper, the DC electric field screening in Si-SiO$_2$ MQW is studied by DC-field induced SHG.
 The magnitude of parameter $\left( {\varepsilon _F/\varepsilon _C} \right)_\bot $ is smaller
than $10^{-3}$ for the structures studied. The experimental bias
dependence of the SHG intensity for MQW reveals the essentially
non-monotonic, oscillatory-like behavior. This is attributed to
the essential quantum effects in the highly anisotropic layered
system with a Coulomb interaction between layers.

\par    MQW structures were fabricated by RF magnetron sputtering on a
vicinal n-Si(100) wafer\cite{14}. MQW structures consist of 40
double amorphous layers of Si-SiO$_2$. Two MQW samples were
studied with SiO$_2$ layers 3 and 5 nm thick. The Si layers were
1.1 nm thick for both samples. The perfect periodic layered
structure of the MQW is studied by X-ray diffraction and on-line
Auger-electron spectroscopy. The amorphous structure of the Si
layers is confirmed by Raman spectroscopy\cite{14}.

\par    The output of a Q-switched YAG:Nd$^{3+}$ laser at a wavelength of 1064-nm was used as fundamental
radiation. A pulse duration of 15 ns and an intensity of 1-10
MW/cm$^2$ avoid laser damage of the sample and other undesired
photo-induced effects. The experimental setup is described in
detail elsewhere\cite{15}. The transparency of the Si substrate at
the fundamental wavelength allows the study of SHG in
transmission. The advantage of this geometry is that only a few
components of the nonlinear susceptibility tensor contribute to
the signal, which simplifies the analysis.

\par    DC-electric-field is applied to the MQW by imposing a bias voltage between
the Al backside electrode and the In-Ga ring gate electrode on the
top of MQW. The free carriers easily move in the in-plane
direction, while the QW's are well isolated from each other,
therefore the voltage distribution inside the structure does not
depend on the in-plane coordinates, although the cap electrode is
not planar.

\par    The in-plane symmetry of the samples was analyzed by measuring the azimuthal SHG
anisotropy. Figure 1 shows the dependence of the SHG intensity on
the azimuthal angle $\psi$. The two-fold symmetry of the
dependence indicates the in-plane anisotropy of the sample
attributed to the initial miscut of the substrate from the (100)
direction. From the microscopic viewpoint, this means the
"terrace" structure of the interface. We expect that these
terraces are reflected in the morphology of several Si and SiO$_2$
layers, closest to the substrate, which possess the $2/m$
point-group symmetry (Figure 1). The deposition of further layers
erodes the anisotropy. Therefore the symmetry of the rest of
layers is expected to be $\infty /m$.

\par    Since both $2/m$ and $\infty /m$ layers possess an inversion symmetry, $P^{2\omega}$ is given by
Eq.(\ref{eq1}). Symmetry of the layer governs the selection rules
for the components of the fourth-rank tensors $\chi ^{(2Q)}$ and
$\chi ^{(3D)}$. Take the $y$-axis to be parallel to the terraces,
and the $z$-axis normal to the surface. For the transmission
geometry $xzxx$, $xzyy$, and $yzxy=yzyx$ components of $\chi
^{(2Q)}$ and $\chi ^{(3D)}$ contribute to the SHG intensity for
the $2/m$ layers, and {\it there is no contribution from $\infty
/m$  layers}. Consequently, the DC-electric-field effects in the
buried Si-SiO$_2$ layers near the vicinal Si substrate are purely
detected.

\par    Figure 2 shows the  dependence of the SHG intensity on the applied bias measured at the maximum
of the rotational anisotropy. Clear oscillatory-like behavior is observed. According to Eq.(\ref{eq1}) this is
related to the oscillatory dependence of the DC-electric-field on the bias.

\par    The following model is developed to obtain the DC-field distribution. First,
we suppose that there is no charge coupling between the upper Si
layer and the metal electrode. Therefore, the current through the
structure is zero, the Fermi level is the same for all QW's and
the Si substrate. Second, since $\left( {\varepsilon
_F/\varepsilon _C} \right)_\bot << 10^{-3}$
 the structure should be considered as a set of charged QW's, with the 2D Fermi-gas of electrons in
each well.

\par     We consider the dependence of $E^{DC}$ and the electrostatic potential $\varphi$
in the middle of the Si layer and the 2D charge density $n$ as
functions of the discrete variable the number of well $l$. The
following discrete analogue of the electrostatic
equations with appropriate boundary conditions is used:

\begin{equation}
\left\{ {\matrix{{-(E^{DC}(l)-E^{DC}(l-1))=-2\pi
(n(l)+n(l-1))/\varepsilon }\cr {\varphi (l-1)-\varphi
(l)=(\varepsilon d + \varepsilon D _1)(E^{DC}(l)+E^{DC}(l-
1))/2\varepsilon}\cr {\varphi (0)=0,\varphi (L)=U_0}\cr }}
\right., \label{eq2}
\end{equation}
where $d$ and $D$ are the thicknesses of the Si and SiO$_2$ layers, respectively; $\varepsilon $ and
$\varepsilon _1$ are static dielectric constants of these layers, $U_0$ is a bias voltage.


\par    We take into account only the first quantized sub-band of the electron energy spectrum of the quantum
well. The expression for $n(\varphi )$ is obtained by the same
procedure as for the 3D case\cite{16} and is given by:

\begin{equation}
n(\varphi )=-{{m_eTe_e} \over {\pi \hbar ^2}}\ln [1+\exp
((e_e\varphi +\mu )/T)] \\
 +{{m_hTe_h} \over {\pi \hbar
^2}}\ln [1+\exp (-(e_h\varphi -\mu +\Delta )/T)], \label{eq3}
\end{equation}
where $\mu$ is the Fermi energy, $\Delta$ is the bandgap, $T$ is
temperature, $m$ is the effective mass of the carrier,
subscripts $e$ and $h$ stand for electrons and holes.

\par    In the limit of $T=0$  and $\Delta=0$ the
system (2) is linear and $\varphi(l)$ takes the form:
\begin{equation}
\varphi (l)=C_1 A^{l}+C_2 A^{-l}, \label{eq4}
\end{equation}
where $C_1$ and $C_2$ are constants and
\begin{equation}
A =\left( 1- \sqrt { \left(\varepsilon d+\varepsilon_1 D \right) /
\varepsilon^2 a_0} \right)\left( 1+\sqrt { \left(\varepsilon
d+\varepsilon_1 D \right) / \varepsilon^2 a_0}\right)^{-1}.
\label{eq5}
\end{equation}
$a_0$ denotes the Bohr radius $\hbar ^2 (me^2)^{-1}$. If
$\left(\varepsilon d+\varepsilon_1 D \right)>\varepsilon^2 a_0$,
then $A$ is negative and $\varphi (l)$ reveals the oscillatory
dependence. This type of dependence remains for the case of the
finite magnitude of bandgap and temperature, as Eqs. (2) should be
solved numerically. Thereafter the SHG bias dependence can be
obtained straightforwardly. As discussed below, the oscillations
in $\varphi (l)$ result in oscillations in $E^{DC}(U_0)$ for the
buried {\it 2/m} layers, which give rise to the SHG signal,
accordingly in Eq.(1).

\par    The total SHG intensity is composed of the DC-field-induced dipole and field-independent quadrupole
terms (Eq. 1). The relative values of the amplitude and phase of
the quadrupole term as well as the flat-bend potential are used as
adjustable parameters. The best approximation is shown in Fig. 2
by solid curves with arrows indicating the bias interval where
only the first quantized electron sub-band is occupied. Outside
this interval the potential dependence of carrier density
$n(\varphi )$ should be modified with the account of the
occupation of the second sub-band. The shift of the bias region of
the model validity is a consequence of the initial bend banding in
the silicon substrate. For the thicker structure the adjusted
position of the flatband potential is naturally shifted to the
higher voltage\cite{7}.

\par     In discussing the origin of the oscillations, it is useful
to compare the system studied above with the classical plasma. In
the latter the spatial distribution of charge density is
monotonic, and the screening is well described by the kinetic
equation for the electron distribution function. This description
by the distribution function requires averaging of fields over the
elementary volume which includes a large number of particles (the
so-called physically infinitesimal volume). This procedure allows
the replacement of the field acting on a particle by the average
field\cite{17}. Generally, the mean-field potential acting on a
particular carrier is formed by all charges in the system {\it
except the carrier itself}. This exclusion is not important for
the 3D Fermi-liquid, in which electron wave functions are
delocalized in space. Another situation occurs in the MQW
structure where the carriers are confined inside the QW's. This
results in the distinct difference between the acting and average
fields. We consider the particular case when the Fermi energy for
the 2D electron liquid in QW's is much smaller than the DC
potential. This means that the interaction between electrons of
the same QW
 is negligible in comparison with their interaction with other charges of the system.
Consequently, the acting potential for the electrons in {\it l}-th
well is the DC-potential formed by external charges all the wells,
{\it except} the $l$-th. On the other hand, the average potential
is still contributed by all the wells.

\par    Figure 3 demonstrates the appearance of oscillations in  $\varphi(U_0)$ due to the discussed
 effect.
At small bias the occupation of all sub-bands in all QW's is zero (thin solid line).
Consider the  potential $U_{0,1}$, when electrons accumulate in well 1.
The potential which determines the occupation number is purely the DC potential
$\varphi _{el}$ of charges at electrodes from the bias supply (the medium dashed line).
 This is worth noting again that the charges in well 1 do not contribute to this potential.
On the other hand, they do contribute to the complete DC potential
$\varphi$ in the system (medium solid  line). The kink in this
line is determined by the charge of well 1, {\it i.e.} by $\varphi
_{el}$. For the certain layer thickness this kink can result in
the negative $\varphi$  in well 1. Similarly, at higher potential
$U_{0,2}$ holes are accumulated in well 2, which is illustrated by
the solid thick curve. One can note from Figure 3, that $\varphi$
and $E^{DC}$ in buried layers appear to be oscillatory functions
of bias $U_0$ in the considered situation.

\par    In conclusion, the DC-electric-field screening in the Si-SiO$_2$ MQW layered
structure is studied both experimentally and theoretically. The
dependence of DC-electric-field $E^{DC}$ on the applied bias
voltage probed in the deep buried Si wells by
DC-electric-field-induced SHG shows essentially non-monotonic,
oscillatory-like behavior. This behavior which is unexpected for
3D Fermi liquids, indicates the importance of the correlations in
electron subsystem for the strongly anisotropic layered system.
The observed phenomenon is explained using the theoretical
description, which takes into account the strong electron Coulomb
localization in the quantum wells due to the high anisotropy of
the effective mass. Because of this localization the electron
subsystem of MQW differs from the 3D Fermi liquid essentially.

\par    We are thankful to V.N. Golovkina for the assistance in experiment and P.V. Elyutin and A.A. Nikulin
for useful discussions. This work was supported by RFBR grants
97-02-17919, 97-02-17923, Special grants for Leading Russian Scientific Schools
96-15-96420 and 96-15-96476, RFBR-DFG grant 98-02-04092,
INTAS-93 grant 0370(ext), INTAS grants YSF-98-9, YSF-98-10, ISSEP grant d98-701, Programs "Center of
Fundamental Optics and  Spectroscopy" and
"Physics of Solid Nanostructures".

{\bf Figure Captions}

Fig. 1. Upper panel: the schematic of MQW structure with vicinal
substrate, and the configuration of the experiment. Lower panel:
Azimuthal dependences of the SHG intensity $I^{2\omega }(\psi )$
for MQW with $d=1.1\,nm$ and $D=5\, nm$, measured in transmission
geometry for parallel polarizations of the fundamental and SH
waves, for two biases: $(\bullet )$ - +2.6 V, $(\circ )$ - +1.26
V. Solid lines are fits to data by Eq.(1) with components of $\chi
^{(2Q)}$ and $\chi ^{(3D)}$, corresponding to the $2/m$ symmetry,
as adjustable parameters. Upper panel: schematics of the sample
and the geometry of experiment.

Fig. 2. Bias dependences of the SHG intensity for MQW with
$D=3\,nm$ (upper panel) and $D=5\,nm$ (lower panel). Solid lines
are fits by the model. Arrows indicate the region of validity of
the model, which is determined by neglecting the second and upper
sub-bands.

Fig. 3. Lower panel: the schematic of the spatial distribution of
potential $\varphi$ (solid lines) across MQW for three
monotonically increased values of $U_0$:
$U_{0,0},U_{0,1},U_{0,2}$. The second subscript indicates the
number of wells participating in the screening of the external
DC-electric field. Directions and relative values of DC-field in
buried layers $E^{DC}$ corresponding to these values of $U_0$ are
shown by arrows. Dashed lines are $\varphi _{el}$ (see text). The
upper panel shows the theoretical dependence of the
DC-electric-field in buried $2/m$ layers $E^{DC}$ as a function of
the external bias $U_0$ for MQW with $d=1.1\,nm$ and $D=5\,nm$.

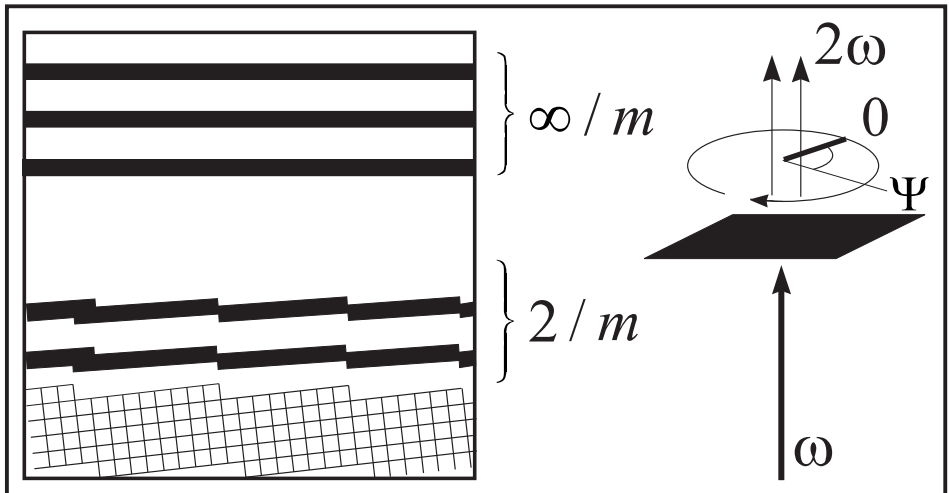
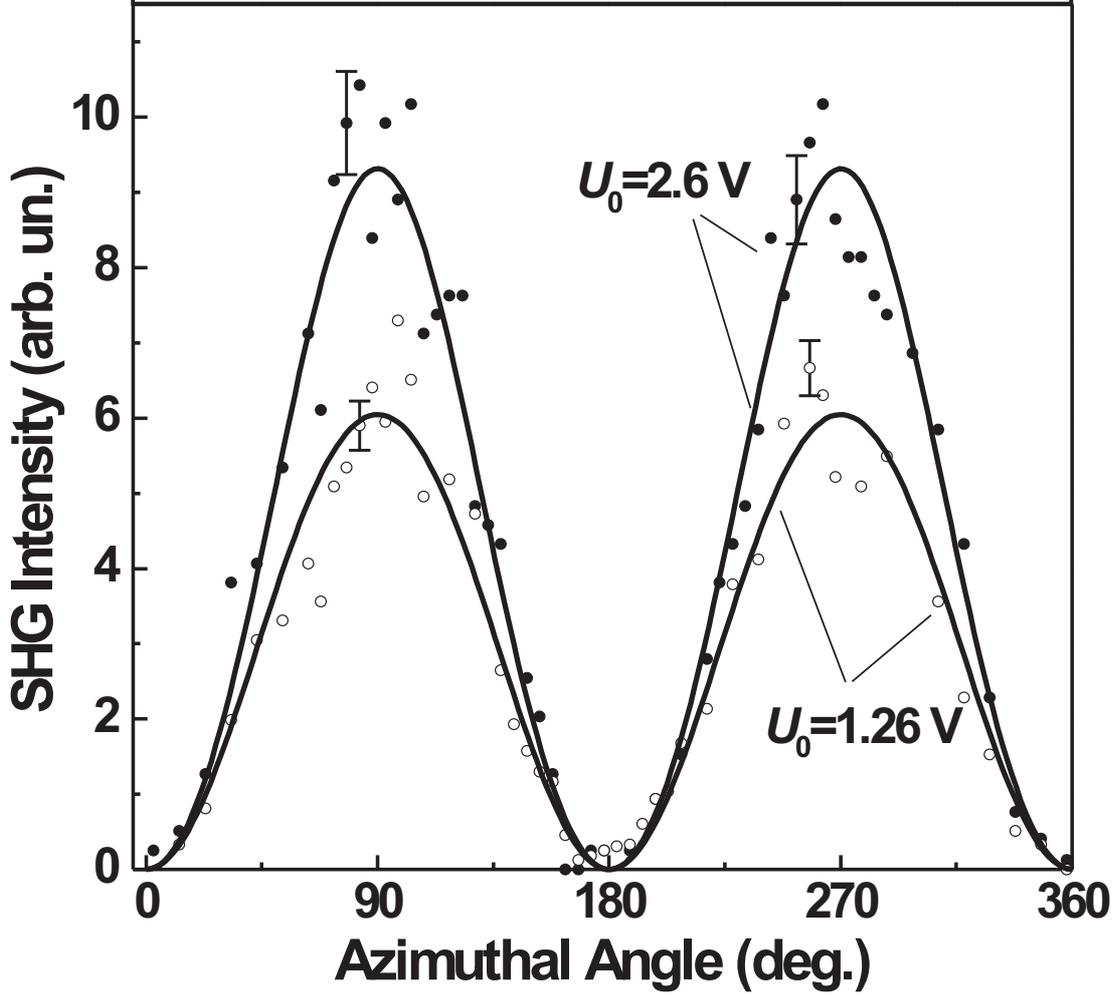
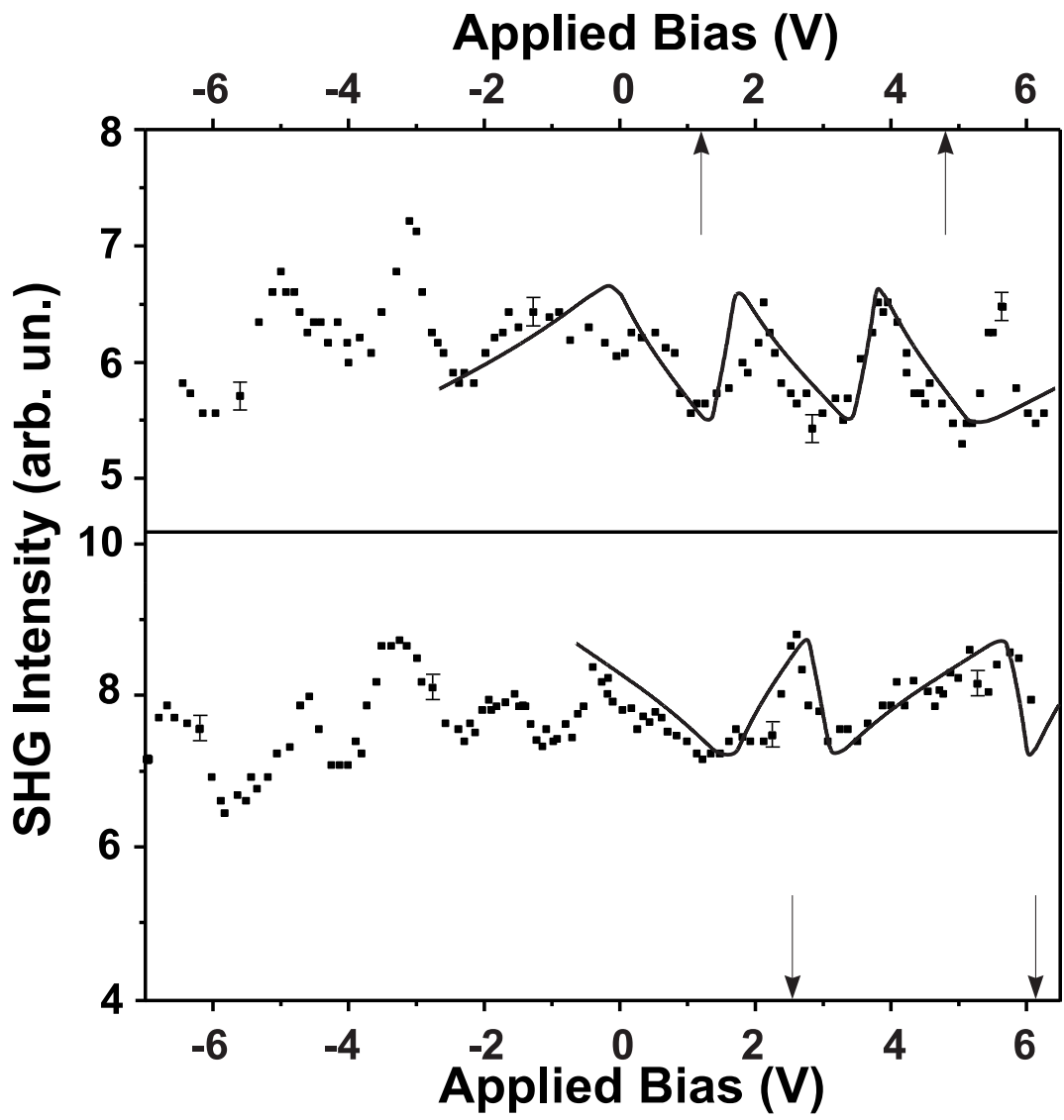
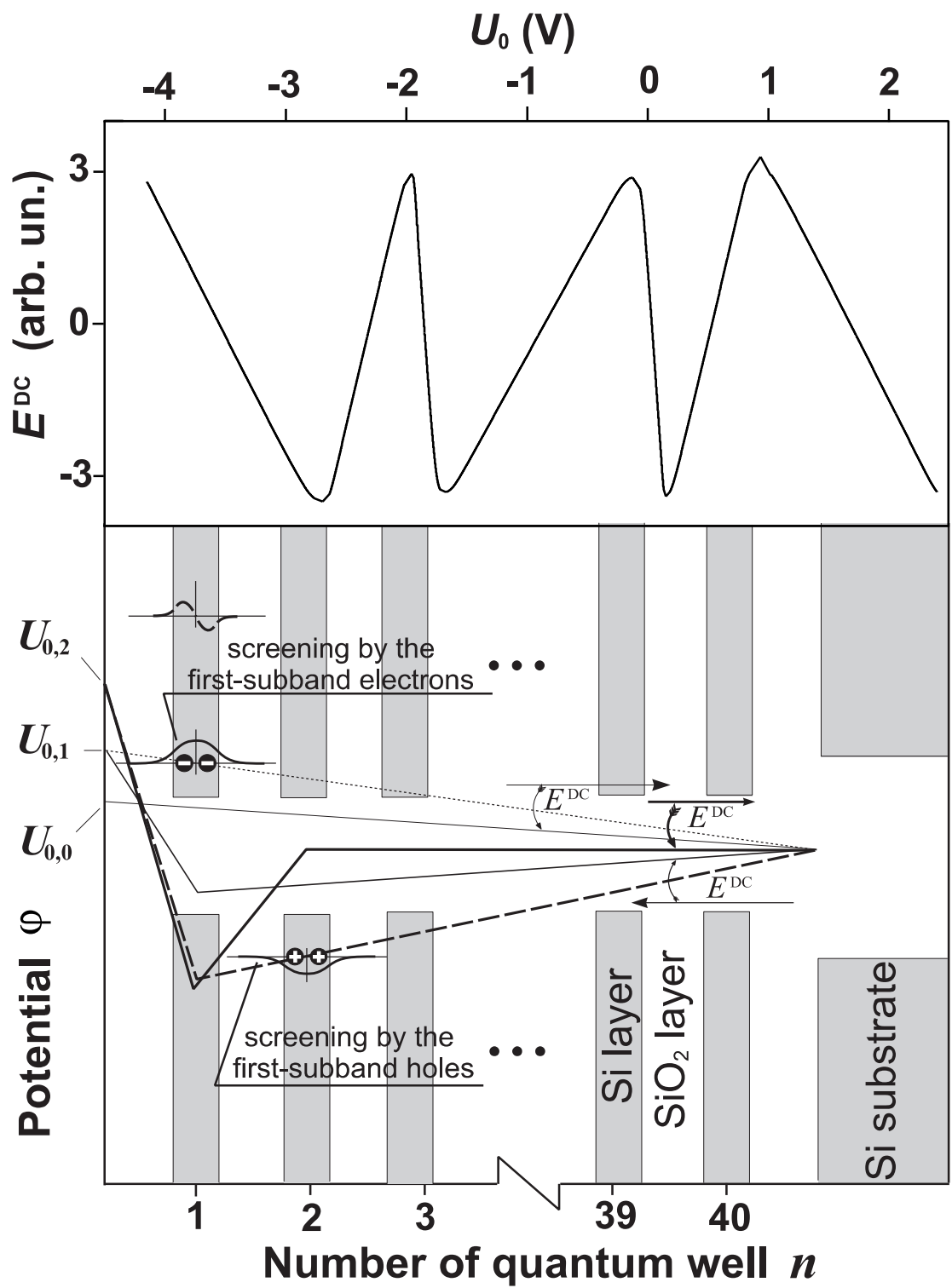
\end{document}